\documentclass[letterpaper]{article} 
\usepackage{aaai24}  
\usepackage{times}  
\usepackage{helvet}  
\usepackage{courier}  
\usepackage[hyphens]{url}  
\usepackage{graphicx} 
\usepackage{natbib}  
\usepackage{caption} 
\usepackage{algorithm}
\usepackage{algorithmic}

\usepackage{newfloat}
\usepackage{listings}
\usepackage{color}

\definecolor{dkgreen}{rgb}{0,0.6,0}
\definecolor{gray}{rgb}{0.5,0.5,0.5}
\definecolor{mauve}{rgb}{0.58,0,0.82}

\DeclareCaptionStyle{ruled}{labelfont=normalfont,labelsep=colon,strut=off} 
\floatstyle{ruled}
\newfloat{listing}{tb}{lst}{}
\floatname{listing}{Listing}
\usepackage{booktabs}
\usepackage{multirow}
\usepackage{adjustbox}
\usepackage{subfigure}
\usepackage{bm}
\usepackage{appendix}
\usepackage{amssymb}
\usepackage{makecell}
\usepackage{amsmath}

\title{Exploring the Potential of Flexible 8-bit Format: Design and Algorithm}
\author{
    Zhuoyi Zhang\textsuperscript{\rm 1}\equalcontrib,
    Yunchen Zhang\textsuperscript{\rm 1,2 }\equalcontrib,
    Gonglei Shi\textsuperscript{\rm 1}\equalcontrib,
    Yu Shen\textsuperscript{\rm 1},
    Ruihao Gong\textsuperscript{\rm 1},
    Xiaoxu Xia\textsuperscript{\rm 1},
    Qi Zhang\textsuperscript{\rm 1},
    Lewei Lu\textsuperscript{\rm 1},
    Xianglong Liu\textsuperscript{\rm 3}
}
\affiliations{
    \textsuperscript{\rm 1}SenseTime Research\\
    \textsuperscript{\rm 2}University of Electronic Science and Technology of China
    \textsuperscript{\rm 3}Beihang University

    \{zhuo1zhang, shigonglei\}@gamil.com, \{zhangyunchen, shenyu, gongruihao, xiaxiaoxu, zhangqi3, luotto\}@sensetime.com, xlliu@nlsde.buaa.edu.cn
}

\usepackage{bibentry}

\begin{document}

\maketitle

\begin{abstract}
   Neural network quantization is widely used to reduce model inference complexity in real-world deployments. However, traditional integer quantization suffers from accuracy degradation when adapting to various dynamic ranges. Recent research has focused on a new 8-bit format, FP8, with hardware support for both training and inference of neural networks but lacks guidance for hardware design. 
   In this paper, we analyze the benefits of using FP8 quantization and provide a comprehensive comparison of FP8 with INT quantization.
   Then we propose a flexible mixed-precision quantization framework that supports various number systems, enabling optimal selection of the most appropriate quantization format for different neural network architectures. 
   Experimental results demonstrate that our proposed framework achieves competitive performance compared to full precision on various tasks, including image classification, object detection, segmentation, and natural language understanding.
   Our work furnishes critical insights into the tangible benefits and feasibility of employing FP8 quantization, paving the way for heightened neural network efficiency in tangible scenarios. 
   Our code is available in the supplementary material.
   
\end{abstract}

\section{Introduction}

Deep neural networks (DNNs) have achieved significant achievements across domains like computer vision, and natural language processing. Nonetheless, their deployment on resource-limited devices poses challenges owing to their computational and memory intensiveness. Several strategies, such as knowledge distillation, sparsity, and quantization, have been developed to mitigate these challenges.

Among these strategies, quantization is a widely adopted technique that involves reducing the precision of the model's weights and activations. The popularity of INT8 quantization~\cite{Jacob_2018_CVPR,2021-neuips-mqbench,li2021brecq,nagel2020up,2020-cvpr-int8training} stems from its ability to deliver impressive performance with minimal accuracy reduction. However, for specific tasks like semantic segmentation and object detection, INT8 quantization can notably impair accuracy~\cite{DBLP:conf/isca/JouppiYAGJKLLMM21}. This has led to the emergence of new data formats, notably BF16 and FP8.

Although hardware support for FP8~\cite{micikevicius2022fp8} is primarily centered around training DNNs, comprehensive exploration, and analysis of FP8 remain in their infancy. While~\cite{huang2021all} did introduce a flexible 8-bit floating-point format, it lacked universality across networks. There's a conspicuous absence of systematic insights regarding the suitability of FP8 or INT8 quantization for various scenarios, and guidance for hardware design is also scarce.



Addressing these gaps, our study provides an in-depth analysis of the pros and cons of diverse data formats and standards. We present a flexible 8-bit design optimized for speed and precision, supporting a blend of different number systems INT and FP to cater to varying data distributions. Importantly, we delineate the data flow of these flexible formats, offering invaluable guidance for hardware design. We've also crafted a highly efficient automatic mixed-precision search algorithm to pinpoint the ideal format combination tailored to distinct network architectures, with E3M4 emerging as dominant across our DNN experiments. Our method's efficacy is underscored by its performance across tasks such as image classification, object detection, semantic segmentation, text-to-image generation, and natural language understanding (NLU), achieving a full-precision level.

To summarize, our contributions are as follows:
\begin{enumerate}
    \item We introduce and select a flexible 8-bit format that is friendly to both speed and precision. Our format supports a mixture of different number systems, including INT and FP, to cope with different data distributions.
    \item We provide an exhaustive outline of the data flow of flexible formats, detailing aspects like data storage and computational processes, thereby steering hardware design toward efficiency.
    \item We have designed an efficient automatic mixed-precision search algorithm to explore the optimal format combination for different network architectures and tasks. Our algorithm can quickly search for the best format for a given network architecture, achieving high performance with low computational cost. 
    \item We demonstrate the effectiveness of our proposed method on various tasks, including image classification, object detection, semantic segmentation, text-to-image generation, and NLU. Our method achieves full-precision level accuracy while being highly efficient and requiring minimal computational resources. 
\end{enumerate}


\section{Related Work}
{\bf Post-training integer quantization.}
Representing FP32 neural networks using fixed-point numbers is prevalent. Since the introduction of the 8-bit integer format quantization by \cite{jacob2018quantization}, it has become a mainstay for model deployment. However, simple 8-bit integer quantization can erode accuracy, particularly given the bell-shaped distribution with long tails often seen in weights or immediate activations of convolutional neural networks. This has led to extensive research aimed at bolstering integer quantization performance. For instance, \cite{banner2019post} suggested a closed-form analytical approach to minimize tensor-level quantization error. \cite{jain2020trained} devised a gradient threshold-based method to even out weight distribution. Techniques to optimize quantization error were explored by \cite{fang2020post}, while \cite{nagel2020up} and~\cite{li2021brecq} innovated weight-rounding mechanisms that cater to input variability.

{\bf Floating-point quantization.}
Float-point quantization, introduced for training by \cite{wang2018training}, has shown efficacy in training DNNs with 8-bit floating point (FP8) numbers across diverse models and datasets. \cite{huang2021all} unveiled a flexible 8-bit floating-point (FFP8) format but overlooked potential hardware implementation costs. Meanwhile, \cite{micikevicius2022fp8} delved into the use of FP8 format encodings for varied tasks, although they omitted some format variations. \cite{kuzmin2022fp} provided a deeper analysis of FP8 formats without giving due consideration to hardware compatibility and integration with integer formats.

{\bf Quantization of language models (LMs).} 
Quantization techniques for LMs have recently garnered significant interest. \cite{DBLP:conf/emnlp/BondarenkoNB21} highlighted the impact of activation outliers in LMs, noting that they can adversely influence quantization precision. Several studies \cite{DBLP:conf/nips/WeiZZGZZYL22, DBLP:journals/corr/abs-2211-10438, DBLP:journals/corr/abs-2304-09145} have shifted the challenges of quantizing activations to weights. \cite{DBLP:conf/nips/YaoAZWLH22} employed per-token quantization for activation and leveraged knowledge distillation to maintain model performance. Furthermore, another approach \cite{DBLP:journals/corr/abs-2208-07339} retains the outlier dimension in activation using FP16 to ensure the model's accuracy.  With LMs’ vastness, weight quantization has emerged as a size-reducing strategy \cite{DBLP:journals/corr/abs-2306-02272, DBLP:journals/corr/abs-2210-17323}. FP quantization, blending precision, and efficiency offer potential future neural network optimization advantages.

\begin{table*}[ht!]
    \begin{center}
    \begin{adjustbox}{max width=\textwidth}
    \begin{tabular}{lllllll}
        \toprule
        \multirow{2}{*}{\bf Format} & {\bf E4M3} & {\bf E4M3} & {\bf E4M3} & {\bf E5M2} & {\bf E5M2} & {\bf E5M2} \\
         & (IEEE 754) & (NIA) & (Ours) & (IEEE 754) & (NIA) & (Ours) \\
        \midrule
        Exponent bias & 7 & 7 & 7 & 15 & 15 & 15\\
        Inf & $S.1111.000_2$ & N/A & N/A & $S.11111.00_2$ & $S.11111.00_2$ & N/A \\
        NaN & 
        \begin{tabular}[l]{@{}l@{}} $S.1111.\{001,$ \\ $ 010, \ldots, 111\}_2$ \end{tabular} 
        & $S.1111.111_2$ & N/A & \begin{tabular}[l]{@{}l@{}} $S.11111.\{01, 10, 11\}_2$ \end{tabular} & $S.11111.\{01, 10, 11\}_2$ & N/A \\
        Zeros & $S.0000.000$ & $S.0000.000$ & $S.0000.000$ & $S.00000.00$ & $S.00000.00$ & $S.00000.00$ \\
        Max normal & $ S.1110.111_2 $ & $ S.1111.110_2 $ & $ S.1110.111_2 $ & $ S.11110.11_2 $ & $ S.11110.11_2 $ & $ S.11110.11_2 $ \\
        Min normal & $ S.0001.000_2 $ & $ S.0001.000_2 $ & $ S.0001.000_2 $ & $ S.00001.00_2 $ & $ S.00001.00_2 $ & $ S.00001.00_2 $ \\
        Max subnormal & $ S.0000.111_2 $ & $ S.0000.111_2 $ & $ S.0000.111_2 $ & $ S.00000.11_2 $ & $ S.00000.11_2 $ & $ S.00000.11_2 $ \\
        Min subnormal & $ S.0000.001_2 $ & $ S.0000.001_2 $ & $ S.0000.001_2 $ & $ S.00000.01_2 $ & $ S.00000.01_2 $ & $ S.00000.01_2 $ \\
        \bottomrule
    \end{tabular}
    \end{adjustbox}
    \end{center}
    \caption{The binary representation of different FP8 formats.}
    \label{tab:different_fp8_formats}
\end{table*}
\section{Preliminaries}
\label{sec:preliminaries}
{\bf Floating-point (FP) Number.} A normal floating-point number $ x \in \mathbb{R} $ is defined as follows:

\begin{equation}
    x = (-1)^s 2^{p-b}(1+\sum_{i=1}^{m} \frac{d_i}{2^i}),
\end{equation}

where $ s \in \{0, 1\} $ represents the sign (0 for positive, 1 for negative), 
$ p $ represents the \textit{e}-bit exponent $ \left( p \in \mathbb{N^+}, 0 < p < 2^e - 1 \right) $, 
and $ b $ is the integer bias number, usually defined as $ 2^{e-1} - 1 $.  
The $ d_i \in \{0, 1\} $ is the \textit{m}-bit mantissa.

To represent the value close to 0, the exponent value \textit{p} = 0 is exclusively allocated for denoting subnormal numbers.
Following IEEE-754, the subnormal number's bias $ b' = b - 1 $. In such case, $ x $ is defined as:

\begin{equation}
    x = (-1)^s 2^{1-b} \sum_{i=1}^{m} \frac{d_i}{2^i}
    \label{eq:sub_num_system}
\end{equation}

{\bf Integer Quantization} maps a floating-point number $ x $ into integer number $ \bar{x} $, which can be explained as:

\begin{equation}
    \label{eq:int_quantization}
    \bar{x} = \mathop{clip}(\lceil \frac{x}{s} \rfloor, -c, c), c = 2^b - 1
\end{equation}

where $ s $ is the quantization step size, $ b $ is the bit setting, $ c $ is the maximum range, and $ \lceil \cdot \rfloor $ indicates the rounding to the nearest function.

\section{Hardware Design For Flexible 8-bit Formats}
\label{sec:hardware_design_for_flexible_formats}

In this section, we present our design guideline for flexible 8-bit formats. 
Initially, we analyze the essential characteristics of subnormal numbers and consider additional formats of FP8 that can be accommodated within a limited chip area. Subsequently, we delve into the mixture of 8-bit formats, including INT and FP, and examine the necessary to enable mixed-format quantization. Finally, we design a reusable and efficient operator for mixed-format inference.

\subsection{The Essential of Subnormal Numbers}
Most of the FP8 formats are designed according to the IEEE-754 standard. Nevertheless, there are also formats, such as the NIA(Nvidia-Intel-Arm) designed format~\cite{micikevicius2022fp8}, which removes certain special number systems to expand the range of representation. Our research indicates that the primary challenge with FP8 quantization is often insufficient precision in characterizing values close to 0, rather than a limitation of the dynamic range of representation. 

For instance, the E2M5 format has a maximum absolute normal value in IEEE-754 format of $ S.10.11111_2 = 1.96875 * 2^1 = 3.9375 $ and a minimum absolute normal value of $ S.01.00000 = 2^0 = 1$. Consequently, any value from -1 to 1 will be truncated to 0. Moreover, it is arduous to scale the magnitude of original data to the range of [1, 3.9375], leading to inadequate quantization accuracy, particularly when using the FP8 format with lower exponents. Therefore, we introduced subnormal values into our number system to increase the representation capability of values in the range of (-1, 1). This allowed E2M5 to represent the minimum absolute value of $ S.00.00001_2 = 2^{-5} = 0.03125 $, according to Eq.~\ref{eq:sub_num_system}.

\subsection{Expanding FP8 Formats within Constrained Chip Areas}

While recent literature \cite{huang2021all,kuzmin2022fp} has explored the representation of FP8 numbers, the hardware design implications, especially for AI inference chips, remain under-discussed. Supporting FP8 numbers introduces hardware design complexities. Specifically, while accommodating special number systems like NaN or subnormal numbers can enhance numerical precision, it may enlarge the register transfer level (RTL) code size and consequently, the overall hardware footprint. Designers are thus tasked with optimizing between special number system adoption and resource allocation.

In our approach, inspired by the IEEE-754 standards and hardware design considerations, we introduce four novel FP8 formats: E5M2, E4M3, E3M4, and E2M5. We contrast the IEEE-754, NIA, and our implementation for the E5M2 and E4M3 formats in Table~\ref{tab:different_fp8_formats}. A comprehensive breakdown of our FP8 system is available in \ref{appendix:detailed_binary_formats}. Distinctively, our approach omits representations of infinities and NaN, which trims the RTL code size. By integrating subnormal numbers, we enhance number representation for magnitudes between 0 and 1. We also furnish two supplementary FP8 formats, granting users the flexibility to better manage diverse tensor distributions. Notably, our FP8 implementation results in a modest hardware area surge of less than 5\% relative to conventional systems.

\subsection{Hardware Support for 8-bit Mixed Format}
\label{sec:hardward}

\begin{figure*}[h!]
    \centering
    \subfigure[]{
        \label{fig:int_mul}
        \includegraphics[width=0.45\linewidth]{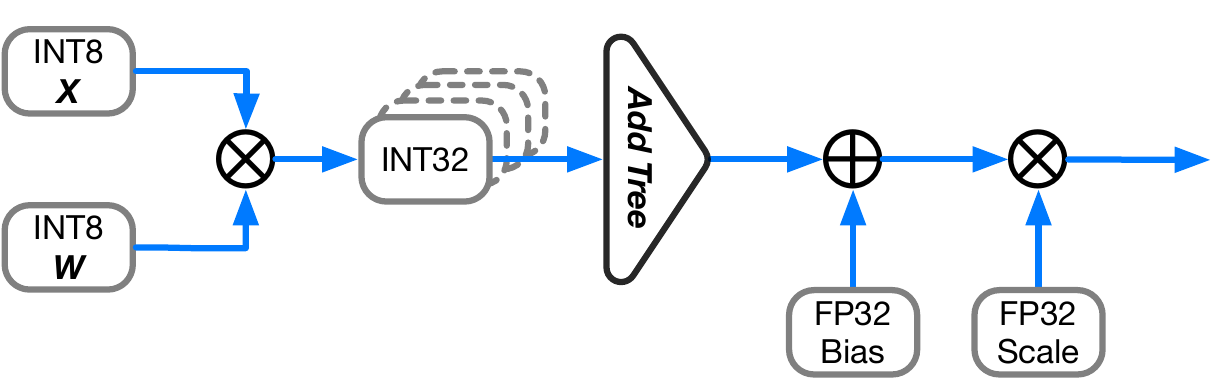}
    }
    \subfigure[]{
        \label{fig:fp_mul}
        \includegraphics[width=0.45\linewidth]{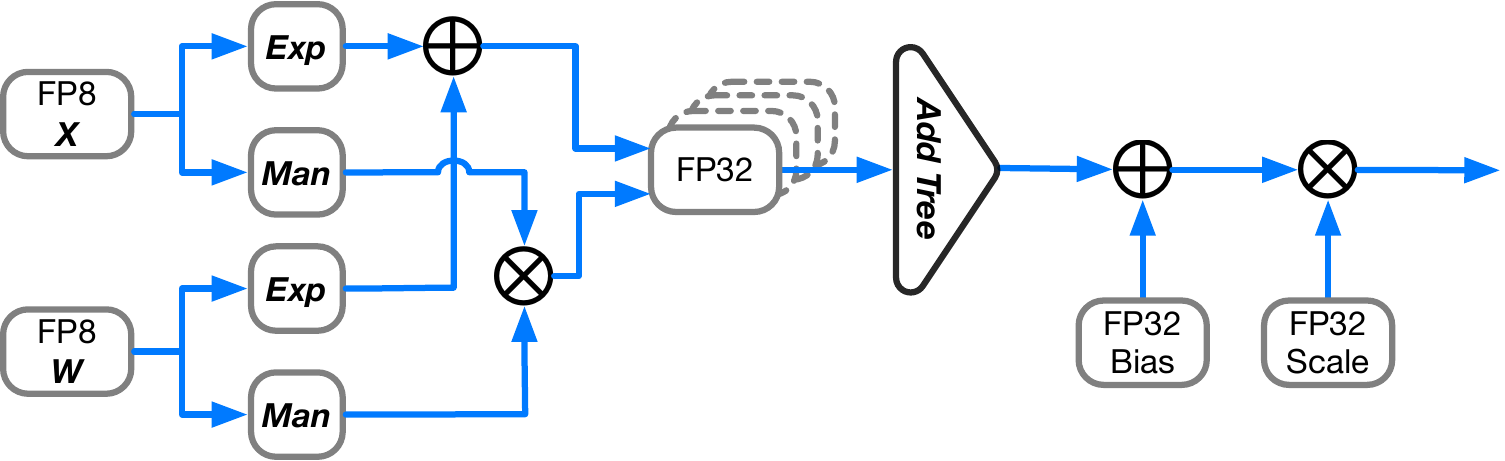}
    }
    \caption{Data flow diagrams for dot product operations in INT8 and FP8 formats. (a) shows the specific flow of dot product between INT8; (b) shows the specific flow of dot product between FP8.}
    \label{fig:int_and_fp}
\end{figure*}


Leveraging a blend of INT8 and FP8 formats is vital for tailoring precision in neural networks, accommodating distinct dynamic ranges and data distributions. To efficiently deploy this mixed-format quantization, hardware compatibility is essential, influencing both system efficiency and performance. We evaluate three schemas: dot products involving solely INT8 or FP8 numbers, and those that mix INT8 and FP8 numbers.

{\bf Dot product between INT8 numbers.} As illustrated in Figure~\ref{fig:int_mul}, INT8 multiplication requires an 8-bit multiplier in the multiplication stage. The resulting product is stored as an INT32, accumulated with other multiplication results. Then, it is added to the bias and multiplied by the FP32 scales to get the final result.

{\bf Dot product between FP8 numbers.} 
As shown in Figure~\ref{fig:fp_mul}, FP8 multiplication requires adding the exponents and multiplying the mantissas to yield an FP32 result. This is subsequently accumulated with other products, added to an FP32 bias, and then scaled for conversion to different number systems. Hardware supporting our FP8 format necessitates 5-bit adders for exponent addition and multipliers for mantissa multiplication. If hardware also accommodates the FP16 system, the 5-bit adder can be repurposed from the FP16 (E5M10) unit. The primary distinction between FP8 and INT8 computations is the 5-bit multipliers and adders in FP8, as opposed to the 8-bit multipliers in INT8, making the power usage for both dot products comparable.

{\bf Dot product between INT8 and FP8.} Our hardware does not support this schema, since it requires an 8-bit multiplier and 5-bit adder. This operation has a higher computational cost compared to multiplying either number system alone. Additionally, there is no significant degradation in accuracy, as demonstrated in \textbf{Section 6.2}, which claims the recommendation to use either the INT8 or FP8 number system for both input and weight data.

\subsection{Detailed Data Flow of Flexible Formats}

\begin{figure}[h!]
    \centering 
    \includegraphics[width=\linewidth]{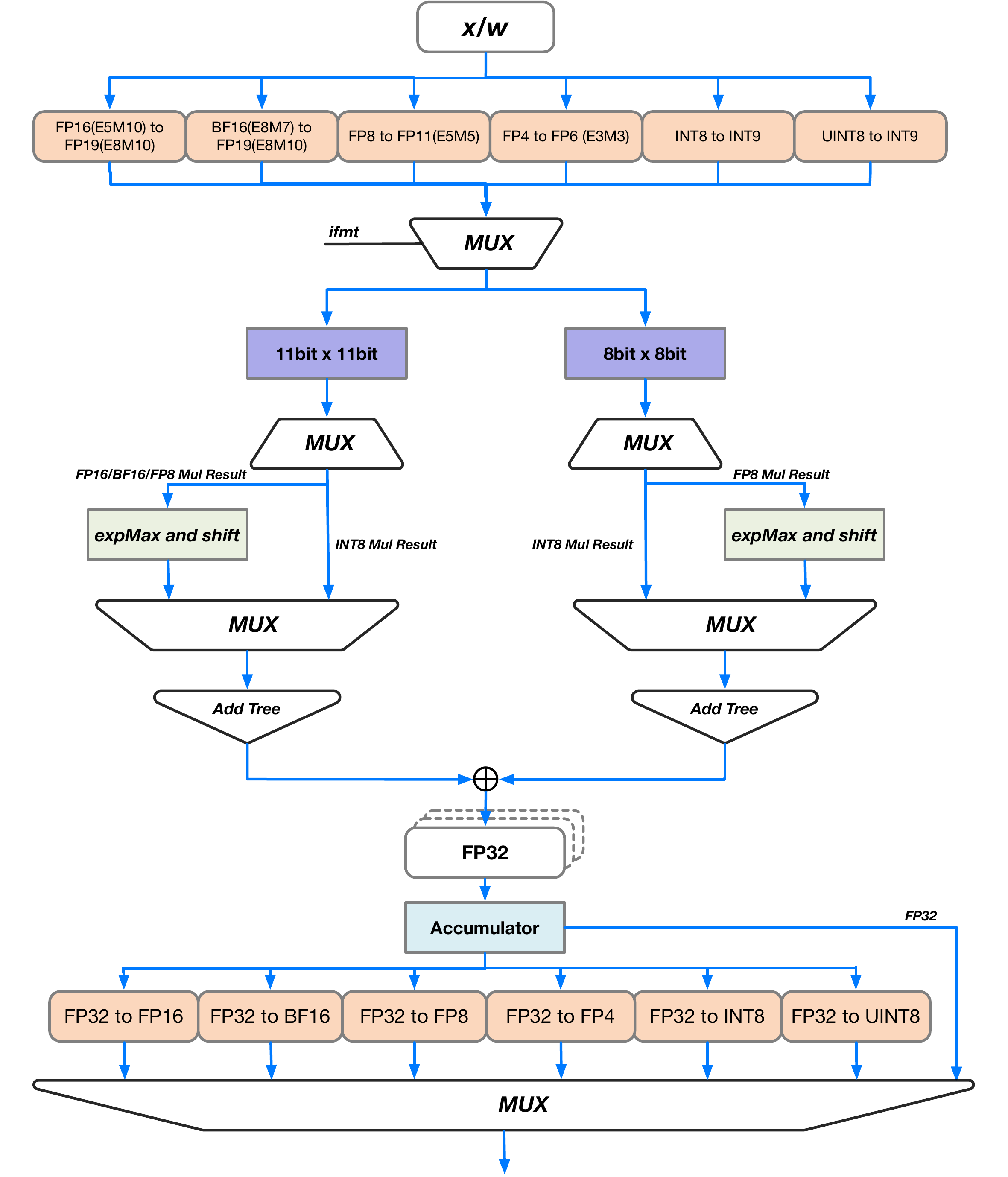}
    \caption{Illustrator of detailed data flow of flexible formats for dot product operations.}
    \label{fig:data_flow}
\end{figure}

The hardware support for flexible formats (including FP16/BF16) is achieved by implementing both INT and FP operations, as well as shared data format schemes. 
The detailed data flow of this process is shown in Figure~\ref{fig:data_flow}.


In the first step of the process, the input data is decoded and converted into three formats of the multiplier, namely FP19 compatible with BF16 and FP16, FP11 compatible with E5M2, E4M3, E3M4, and E2M5, and INT9 compatible with INT8 and UINT8, as shown in Figure~\ref{fig:data_flow}. 
The input data is then split into two streams and sent to two groups of multipliers for multiplication. The left path is used for multiplying FP19 with other inputs based on the decoding result. 
Additionally, in the case of INT9 input, both paths can be used for multiplication, which increases the computing power of INT8/UINT8 compared to FP16/BF16.

For FP11 input, it can be calculated in both paths similar to INT9, thus making the computing power of the FP8 number almost equivalent to that of INT8/UIN8. The multiplication results are then sent to the add tree, and for INT9/FP11, both streams can be calculated for their results, while for FP19, only one stream can be calculated while the other is leisure. After the results from the two streams are added, they are converted to the output number system of the accumulator. The implementation of these steps is crucial for achieving hardware support for flexible formats.


\section{Fast and Automatic Format Selection}


This section aims to address the challenge of selecting an optimal mixed format for neural network inference in a fast and automatic manner. Firstly, we introduce a unified formulation for both INT and FP quantization, which enables consistent treatment of the two formats. Then, we propose a resolution-aware strategy that considers the dynamic range of input data and is quantization-agnostic for efficient format selection. This approach allows for the automatic selection of the best format for each layer of the neural network without manual intervention. Additionally, to facilitate FP8 simulation, we provide CUDA kernels for expedient experimentation.

\subsection{Unified INT and FP Quantization Formulation}
\label{sec:unified_formulation}

In order to enhance reusability, we propose a unified quantization formulation for both INT and FP that utilizes the concept of resolution $ r $. The formulation can be expressed as follows:

\begin{equation}
    \bar{x} = \mathop{clip} (\lceil \frac{x}{r} \rfloor, -c, c),
\end{equation}

where $ r = 2^{\mathop{log}_2 \lfloor x \rfloor -m} $ is calculated based on the number of mantissa and the value of the exponent, and $ c = 2^m -1 $ is used for FP quantization. For INT quantization, $ r = s $ and $ c = 2^b - 1 $ as described in Eq.~\ref{eq:int_quantization}. 
This unified quantization formulation enables the hardware to reuse many modules and algorithms that are initially intended for INT quantization. Therefore, this approach provides greater flexibility in hardware implementation and makes quantization algorithms plug-and-play.

\subsection{Resolution-aware Optimal Format for Quantization}
\label{sec:resolution-aware}

The quantization error $ (\Delta) $ is typically defined as the mean squared error (MSE) between the original input $ \bm{X} $ and the quantized output $ Q(\bm{X}) $, which comprises two types of errors: clipping error $ \Delta_{clip} $ and the rounding error $ \Delta_{round} $, expressed by:

\begin{equation}
    \begin{aligned}
        \Delta &= \Delta_{clip} + \Delta_{round} \\
               &= \frac{1}{I} \sum_{\vert x_i \vert > c} (\vert x_i \vert -c)^2 + \frac{1}{I} \sum_{\vert x_i \vert \leq c} (x_i - x_i^q)^2,
    \end{aligned}
    \label{eq:mse}
\end{equation}

where $ I $ is the total number of elements in the input tensor $ \mathbf{X} $, $ x_i $ and $ x_i^q $ are the $ i $-th element of $ \mathbf{X} $ and $ Q(\mathbf{X}) $, respectively. Omitting $ \Delta_{clip} $, integer quantization assigns the same resolution to all elements, whereas floating-point quantization provides different resolutions, based on the value of the element exponent and $ m $, as mentioned in \textbf{Section 5.1}. In terms of resolution, $ \Delta $ can be rewritten as:

\begin{equation}
    \begin{aligned}
        \Delta &=  \frac{1}{I} \sum_{\vert x_i \vert \leq c} r_i^2 (\frac{x_i}{r_i}  - \lceil \frac{x_i}{r_i} \rfloor )^2 \leq \frac{1}{4I} \sum_{\vert x_i \vert \leq c} r_i^2,
    \end{aligned}
    \label{eq:avg_res}
\end{equation}

where $ r_i $ denotes the resolution of $ i $-th element. The average resolution determines the level of granularity in quantization. For a given tensor, we can use the average resolution given in Eq.~\ref{eq:avg_res} to decide which format we choose. This quantization-agnostic strategy can adapt to the characteristics of different tensors more easily and efficiently, making it a competitive alternative to calculating the MSE between input and quantized output directly. The specific algorithmic flow is represented in \textbf{Appendix.4}.




\section{Experiments}


In this section, we aim to validate our flexible data format designs for DNNs' inference and the efficiency of the resolution-aware format selection. First, we describe our experimental setup and quantization method in \textbf{Section 6.1}. In \textbf{Section 6.2}, we present our baseline results with different format combinations. In \textbf{Section 6.3}, we discuss the influence of implementation details differences. In \textbf{Section 6.4}, we clarify the results of format selection. More ablation experiments such as comparisons of low-bit and generative models are placed in the Appendix.

\subsection{Setups}
\label{sec:exp_setups}

{\bf Implementation Details.}
For post-training quantization (PTQ), we set up our experiments as follows. We use 256 random calibration samples for models, then utilize a per-tensor and symmetrical method for quantization, obtaining $ s $ using the MinMax method~\cite{DBLP:journals/corr/abs-1806-08342}. Format selection of mixed formats including INT8 is determined by evaluating MSE loss of output among all candidate formats. 

{\bf Baseline:} In order to demonstrate the superiority of the FP8 format that we propose, we conduct extensive experiments on three common tasks in computer vision, using different models. For classification, we evaluate: ResNet-18, ResNet-50~\cite{he2016resnet}, MobileNetv2~\cite{sandler2018MobileNetv2}, RegNet3.2G~\cite{radosavovic2020designing}, EfficientNet-b3~\cite{tan2019efficientnet}, and ViT~\cite{dosovitskiy2020vit} on ImageNet dataset~\cite{deng2009imagenet}. We also evaluate our framework in object detection: Retina-FPN~\cite{DBLP:journals/pami/LinGGHD20}, Faster-RCNN~\cite{ren2015faster}, YOLOX-m, and YOLOX-l~\cite{DBLP:journals/corr/abs-2107-08430} on COCO dataset~\cite{lin2014microsoftCOCO}. For segmentation, we compare the results of: DeepLabv3-R50~\cite{DBLP:journals/corr/ChenPSA17}, PSPNet-R50~\cite{DBLP:conf/cvpr/ZhaoSQWJ17}, UNet-1024~\cite{ronneberger2015u}, and SegFormer-b0~\cite{DBLP:conf/nips/XieWYAAL21} on cityscapes dataset~\cite{cordts2016cityscapes}. For the language model, we test BERT~\cite{DBLP:conf/naacl/DevlinCLT19} on the GLUE benchmark \cite{DBLP:conf/iclr/WangSMHLB19}.


\subsection{Main Results}
\label{sec:main_results}

\begin{table*}[ht!]
    \begin{center}
    \begin{adjustbox}{max width=\textwidth}
    \begin{tabular}{l|cccccccc}
        \toprule
        \multirow{1}{*}{\bf Task (metric)} &{\bf Model} & {\bf FP32} & {\bf INT8} & {\bf NIA Format} & {\bf Mixed FP8} & {\bf Mixed FP8 ({\it r})} & {\bf All Mixed} & {\bf Limited Mix} \\
        \midrule
        \multirow{6}{*}{Classification} & ResNet-18 & 70.28 & 69.73 & 69.16 & 70.17 & 69.77 & 70.13 & 70.17 \\
        \multirow{6}{*}{(Top-1 Acc.)} & ResNet-50 & 76.76 & 76.17 & 76.04 & 76.70 &76.40 & 76.74 & 76.70 \\
        \multirow{6}{*}{} & MobileNetv2 & 73.26 & 68.48 & 65.44 & 72.47 & 71.64 & 72.60 & 72.25 \\
        \multirow{6}{*}{} & RegNet3.2G & 78.43 & 78.07 & 77.61 & 78.18 & 78.17 & 78.29 & 78.29 \\
        \multirow{6}{*}{} & EfficientNet-b3 & 80.19 & 58.49 & 78.71 & 79.98 & 79.93 & 80.04 & 79.99 \\
        \multirow{6}{*}{} & ViT & 83.15 & 82.67 & 82.92 & 83.08 & 83.06 & 83.17 & 83.08 \\
        \midrule
        \multirow{3}{*}{Detection} & Retina-FPN & 37.00 & 36.21 & 35.93 & 36.85 & 36.64 & 36.85 & 36.85 \\
        \multirow{3}{*}{(mAP)} & Faster-RCNN & 38.27 & 37.53 & 37.20 & 38.21 & 37.57 & 38.16 & 38.21 \\
        \multirow{3}{*}{} & YOLOX-m & 44.36 & 43.15 & 43.63 & 44.32 & 43.82 & 44.29 & 44.32 \\
        \multirow{3}{*}{} & YOLOX-l & 47.35 & 45.96 & 46.63 & 47.27 & 46.81 & 47.35 & 47.27 \\
        \midrule
        \multirow{3}{*}{Segmentation} & DeepLabv3-R50 & 79.46 & 72.97 & 79.25 & 79.31 & 78.77 &79.32 & 79.31 \\
        \multirow{3}{*}{(mIoU)} & PSPNet-R50 & 77.46 & 71.29 & 77.27 & 77.15 & 76.94 & 77.28 & 77.13 \\
        \multirow{3}{*}{} & UNet-1024 & 67.40 & 65.28 & 65.62 & 67.33 & 66.93 & 67.31 & 67.28 \\
        \multirow{3}{*}{} & SegFormer-b0 & 75.28 & 74.52 & 74.85 & 75.00 & 75.27 & 75.24 & 75.24 \\
        \bottomrule
    \end{tabular}
    \end{adjustbox}
    \end{center}
    \caption{Precision of various 8-bit quantization formats on classification, detection, and segmentation models. The All Mixed method outperforms other approaches. Mixed FP8 (r) refers to Mixed FP8 with a resolution-aware strategy that enables fast scale and format quantization search, achieving a result close to the best one. Limited Mix, which is hardware-friendly, exhibits minimal accuracy loss.}
    \label{tab:baseline}
\end{table*}

\begin{table*}[ht!]
    \begin{center}
    \begin{adjustbox}{max width=\textwidth}
    \begin{tabular}{l|cccccccc}
        \toprule
        \multirow{1}{*}{\bf Task} & {\bf FP32} & {\bf INT8} & {\bf NIA Format} & {\bf Mixed FP8} & {\bf Mixed FP8 ({\it r})} & {\bf All Mixed} & {\bf Limited Mix} & {\bf W4A8}\\
        \midrule
        CoLA & 59.60 & 54.96 & 58.24 & 58.59 & 58.10 & 58.33 & 58.33 & 56.25 \\
        SST-2 & 93.35 & 91.86 & 93.15 & 93.23 & 93.23 & 93.12 & 93.12 & 92.66  \\
        MRPC & 87.75 & 77.45 & 86.48 & 86.82 & 87.25 & 87.5 & 87.5 & 87.25 \\
        STS-B & 89.70 & 84.80 & 89.69 & 89.72 & 89.69 & 89.65 & 89.65 & 88.62 \\
        QQP & 90.91 & 89.05 & 90.78 & 90.80 & 90.82 & 90.81 & 90.81 & 89.50 \\
        MNLI & 84.94 & 81.18 & 84.90 & 84.94 & 84.93 & 84.95 & 84.95 & 82.31 \\
        MNLI-mm & 84.76 & 81.45 & 84.80 & 84.86 & 84.79 & 84.84 & 84.84 & 81.26 \\
        QNLI & 91.84 & 88.50 & 91.69 & 91.73 & 91.82 & 91.74 & 91.74 & 90.43 \\
        RTE & 72.56 & 67.87 & 71.84 & 72.56 & 72.20 & 72.56 & 72.56 & 70.04 \\
        \midrule
        Avg. & 83.93 & 79.68 & 83.51 & 83.69 & 83.65 & 83.72 & 83.72 & 82.04 \\
        \bottomrule
    \end{tabular}
    \end{adjustbox}
    \end{center}
    \caption{Performance comparison of various quantization formats on the BERT model. While FP8-related formats show similar performance to the FP32 baseline, the aggressive W4A8 quantization yields a respectable average, demonstrating only a 2.2\% decline from the FP32 results.}
    \label{tab:glue}
\end{table*}



We apply the PTQ method to all the models mentioned above, aiming to compare different 8-bit formats or their mixtures fairly. The main results are presented in Table Table~\ref{tab:baseline}. The results in each column are obtained using the same configuration except for the number format.

The experiments are conducted using several quantization formats, including \textbf{FP32}, \textbf{INT8}, \textbf{NIA format}, \textbf{Mixed FP8}, \textbf{Mixed FP8(r)}, \textbf{All Mixed}, and \textbf{Limited Mix}. The \textbf{FP32} format is used to represent the full-precision model. For \textbf{INT8}, we employ the standard 8-bit integer quantization method. \textbf{NIA format} is using a FP8 format co-developed by Nvidia, Intel, and Arm. \textbf{Mixed FP8} is implemented using our four FP8 formats. \textbf{Mixed FP8(r)} represents \textbf{Mixed FP8} with a resolution-aware strategy. \textbf{All Mixed} refers to all weights and activations that can choose any format from INT8, four FP8 formats. {\bf Limited Mix} only allowing weights and activations to use the same number system, INT or FP.
 
{\bf Results on CV tasks.} In Table~\ref{tab:baseline}, we observe that all of our mix-format methods show significant improvement compared to the {\bf INT8} precision. For the classification, detection, and segmentation tasks, {\bf All Mixed} resulted in an average increase of 4.6\%/1.0\%/3.8\% compared to {\bf INT8} and approximately 0.07\%/0\%/0.09\% compared to {\bf Mixed FP8}. It is observed that the INT8 candidate can improve the accuracy slightly. Compared to {\bf INT8}, {\bf Mixed FP8(r)} brings about 4.23\%/0.50\%/3.46\% increase, and {\bf Limited Mix} brings 4.48\%/0.95\%/3.73\% increase on each task on average. Although {\bf Mixed FP8(r)} has a minor loss of accuracy compared to {\bf Mixed FP8}, it achieves a speed-up in quantization speed. {\bf Limited Mix} also has a minor loss of accuracy compared to {\bf All Mixed} but is more hardware-friendly. For \textbf{NIA format}, there appears to be a significant fluctuation. For instance, the relatively large loss of precision that occurs on the MobileNetv2 is even lower than the precision of INT8(-3.04\%). But then almost exceeds \textbf{All Mix} (-0.01\%) on PSPNet. While in general, our format mixing solution shows high performance while also speeding up the search process and improving hardware efficiency with only a minor loss of precision.

{\bf Results on NLP tasks.} 
Table~\ref{tab:glue} presents the precision of various quantization formats on the BERT model. The FP32 baseline achieved an average precision of 83.93\%. When quantized to {\bf INT8}, a decline to 79.68\% was noted. Impressively, all FP8-related formats, including {\bf NIA Format} (83.51\%), {\bf Mixed FP8} (83.69\%), {\bf Mixed FP8 (r)} (83.65\%), and both {\bf All Mixed} and {\bf Limited Mix} (83.72\%), performed consistently close to the FP32 baseline. On the other hand, the aggressive W4A8 format (4-bit weight and 8-bit activation), yielded a respectable average of 82.04\%, marking only a modest 2.2\% drop from the FP32 precision. In summary, while FP8 formats emulate the original performance admirably, the more radical W4A8 approach proves to be promising with minimal accuracy compromises.

\subsection{Ablation Study}
\label{sec:ablation_study}

\begin{table}[ht!]
    \begin{center}
    \begin{adjustbox}{max width=\textwidth}

    \begin{tabular}{lccccc}
        \toprule
        & {\bf E2M5} & {\bf E3M4} & {\bf E4M3} & {\bf E5M2} & {\bf Mix} \\
        \midrule
        Disable-sub & 0.10 & 67.13 & {\bf 76.21} & 73.68 & 76.34 \\
        Enable-sub & {\bf 76.11} & {\bf 76.70} &  76.13 & {\bf 73.68} & {\bf 76.70} \\
        \bottomrule
    \end{tabular}
    
    \end{adjustbox}
    \end{center}
    \caption{The impact of subnormal enabled on the accuracy of ResNet-50 8-bit quantization. Subnormal Enabled shows strong compatibility on different FP8 formats. The original ResNet-50 FP32 accuracy is 76.76, and could be found at Table \ref{tab:baseline}.}
    \label{tab:subnormal}
\end{table}
 
  

\textbf{Incorporating Subnormal Numbers is Crucial.}
We assessed the role of subnormals in FP8 formats by comparing the performance of ResNet-50 with and without subnormal numbers. This evaluation, detailed in Table~\ref{tab:subnormal}, shows that the accuracy variance is particularly pronounced as the number of mantissa bits increases. With subnormals disabled, the overall standard deviation and average accuracy are 29.49\% and 58.69\%, respectively. In contrast, when subnormals are enabled, these figures improve to 1.12\% and 75.86\%. Hence, integrating subnormal numbers bolsters compatibility across varying FP formats.

\textbf{Extending Number Count in NIA Format is Redundant.}
In our implementation of the 8-bit float number format, we opted against extending as suggested in \cite{micikevicius2022fp8}. The maximum normal values for E2M5, E3M4, E4M3, and E5M2 are 3.9375, 15.5, 240, and 57344, respectively. We believe these combinations sufficiently represent numbers in weight and activation. Evidence from Table~\ref{tab:baseline} and Table~\ref{tab:glue} shows that our mixed-precision quantization consistently outperforms the NIA format across all tasks. This suggests that extending numbers in the FP8 format does not significantly enhance accuracy.

 
\setlength\tabcolsep{3.0pt}
\begin{table}[h!]
    \begin{center}
    \begin{adjustbox}{max width=\textwidth}
    \begin{tabular}{lccccc}
    \toprule
        {\bf Model} & {\bf FP32} & {\bf Mixed FP6} & {\bf Mixed FP6 ({\it r})} & {\bf Speed Up} \\
    \midrule
        PSPNet-R50 & 77.46 & 71.15 & \textbf{72.87} & x1.55  \\
        UNet-1024 & 67.40 & 60.47 & \textbf{63.10} & x1.48  \\
    \bottomrule
    \end{tabular}
    \end{adjustbox}
    \end{center}
    \caption{A comparison of the search speed and accuracy between Mixed FP and Mixed FP(r) at 6-bit level on segmentation models.}
    \label{tab:my_label}
\end{table}
{\bf Effect of resolution-aware strategy.} 
In \textbf{Section 5.2}, we propose a method for determining the number format in a more efficient way. Our proposed method similarly affects 8-bit number format choice, as demonstrated in Table~\ref{tab:baseline}. 
Our proposed method achieves better results in low-bit format combinations, as shown in Table~\ref{tab:my_label}. We implement FP6 as E2M3 and E3M2, similar to FP8 (specific details of the FP6 implementation can be found in \ref{appendix:detailed_binary_formats}). {\bf Mixed FP6(r)} achieves a 1.72\% and 2.63\% improvement over {\bf Mixed FP6} on PSPNet and UNet, respectively, and also increases the search speed by 1.5.
In conclusion, our resolution-aware strategy provides a faster search speed and even better results in low-bit format combinations, making it an effective method for quantization in segmentation models.

\textbf{Differential Impacts of FP8 and INT8 on Various Tasks.}
Experimental results highlight that FP8 offers varied performance enhancements over INT8 across different tasks. For instance, while MobileNetv2 witnesses a significant 3.99\% improvement with Mixed FP8 over INT8, the gain for ResNet-50 is a modest 0.53\%. This discrepancy can potentially be attributed to differences in activation ranges. As illustrated in Figure~\ref{fig:his_comp}, MobileNetv2 exhibits a broader and more dispersed parameter distribution. Hence, FP quantization seems to be especially advantageous for models with expansive activation distributions.

\subsection{Identifying the Predominant 8-bit Format}
\label{sec:dominant_format}

\begin{figure}[t]
    \centering
    \subfigure[]{
        \includegraphics[width=0.45\linewidth]{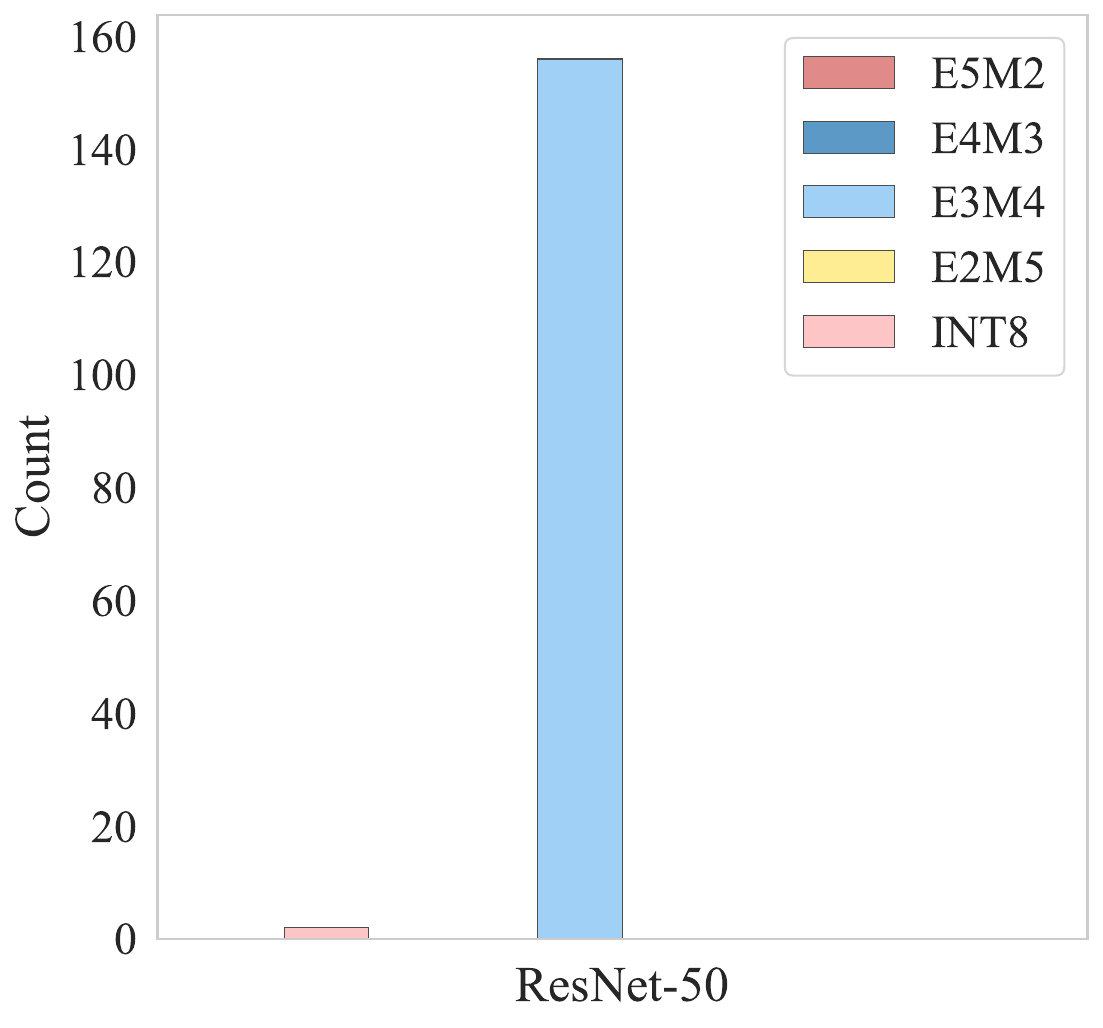}
    }
    \subfigure[]{
        \includegraphics[width=0.45\linewidth]{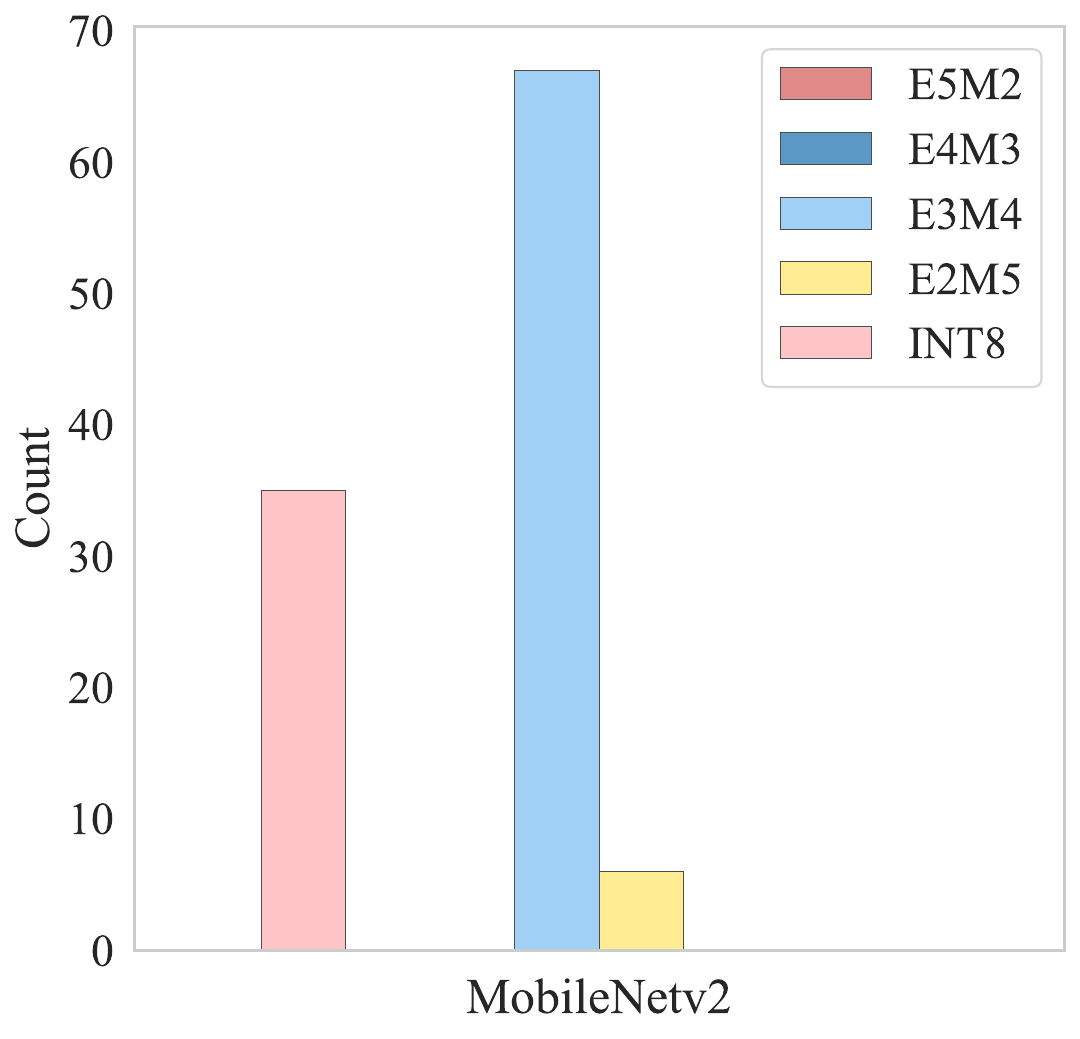}
    }
    \caption{Two patterns of format selection. (a) is the number format selection of weights and activation values of ResNet-50 and (b) is of MobileNetv2.}
    \label{fig:model}
\end{figure}

Our experimental data elucidates number format preferences among various models. Using ResNet-50 and MobileNetv2 as case studies, Figure~\ref{fig:model} depicts two distinct allocation patterns. Predominantly, ResNet-50 favors the E3M4 format, relegating INT8 to a distant second, with other formats being overlooked entirely. MobileNetv2's inclination is also towards E3M4, albeit with INT8 securing roughly half the format allocation. The E2M5 format seldom engages in quantization, and akin patterns emerge across other models, solidifying E3M4's status as the preferred 8-bit format.

\begin{figure}[t]
    \centering
    \includegraphics[width=0.8\linewidth]{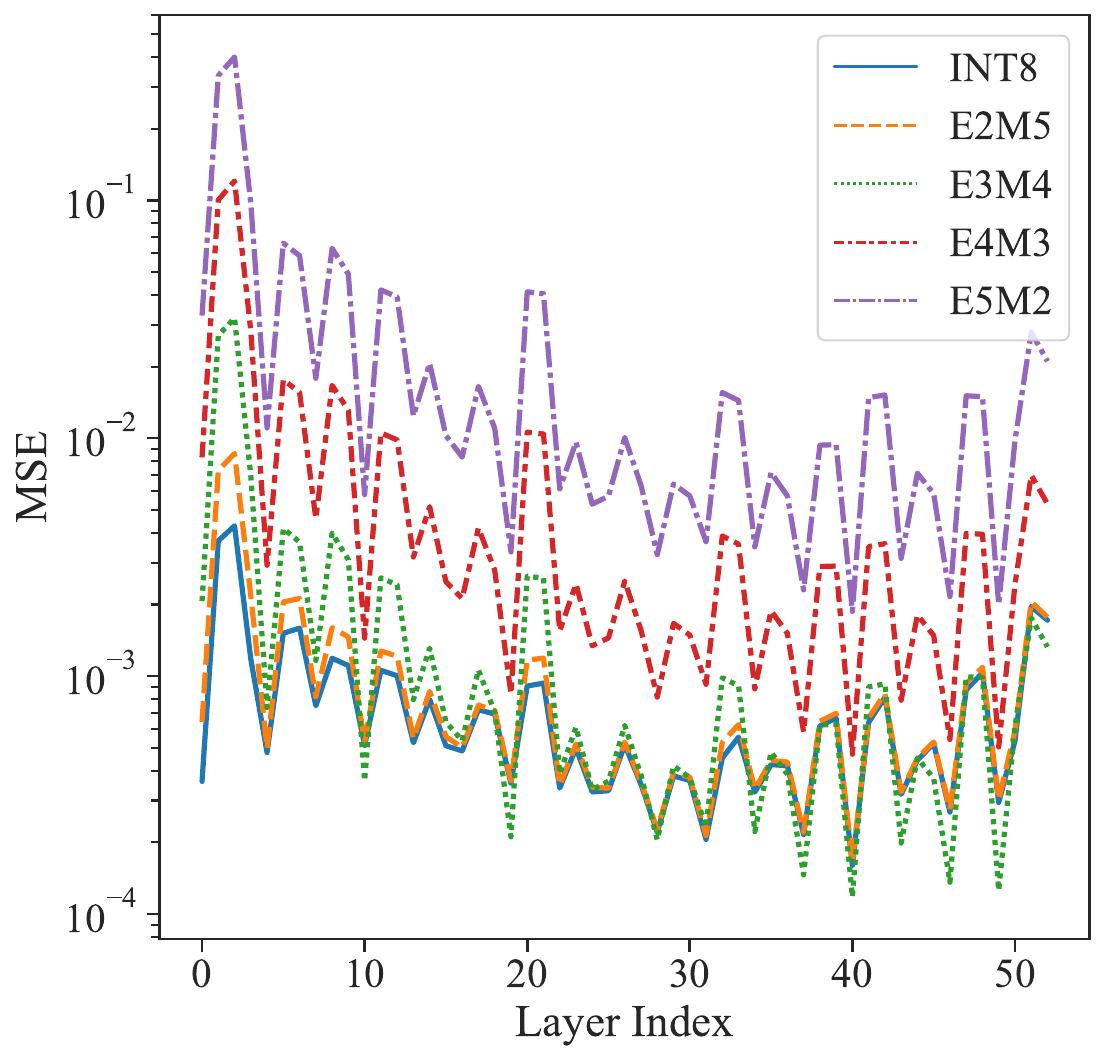}
    \caption{Quantization loss for all layers of MobileNetv2.}
    \label{fig:line}
\end{figure}

Furthermore, the layer's position impinges upon the format selection. Utilizing MobileNetv2 as a reference, we collect the MSE loss in each layer's activations across various formats while leaving weights non-quantized. Figure~\ref{fig:line} reveals E3M4's supremacy in minimizing MSE loss throughout layers, save for the initial 20 that lean towards INT8. E2M5 closely shadows INT8's loss curve, attributed to its homogenous representation stemming from an enriched mantissa. Nonetheless, neither E2M5 nor E3M4 consistently emerges as optimal across layers.

\begin{figure}[t]
    \centering
    \subfigure[]{
        \label{fig:his_comp}
        \includegraphics[width=0.45\linewidth]{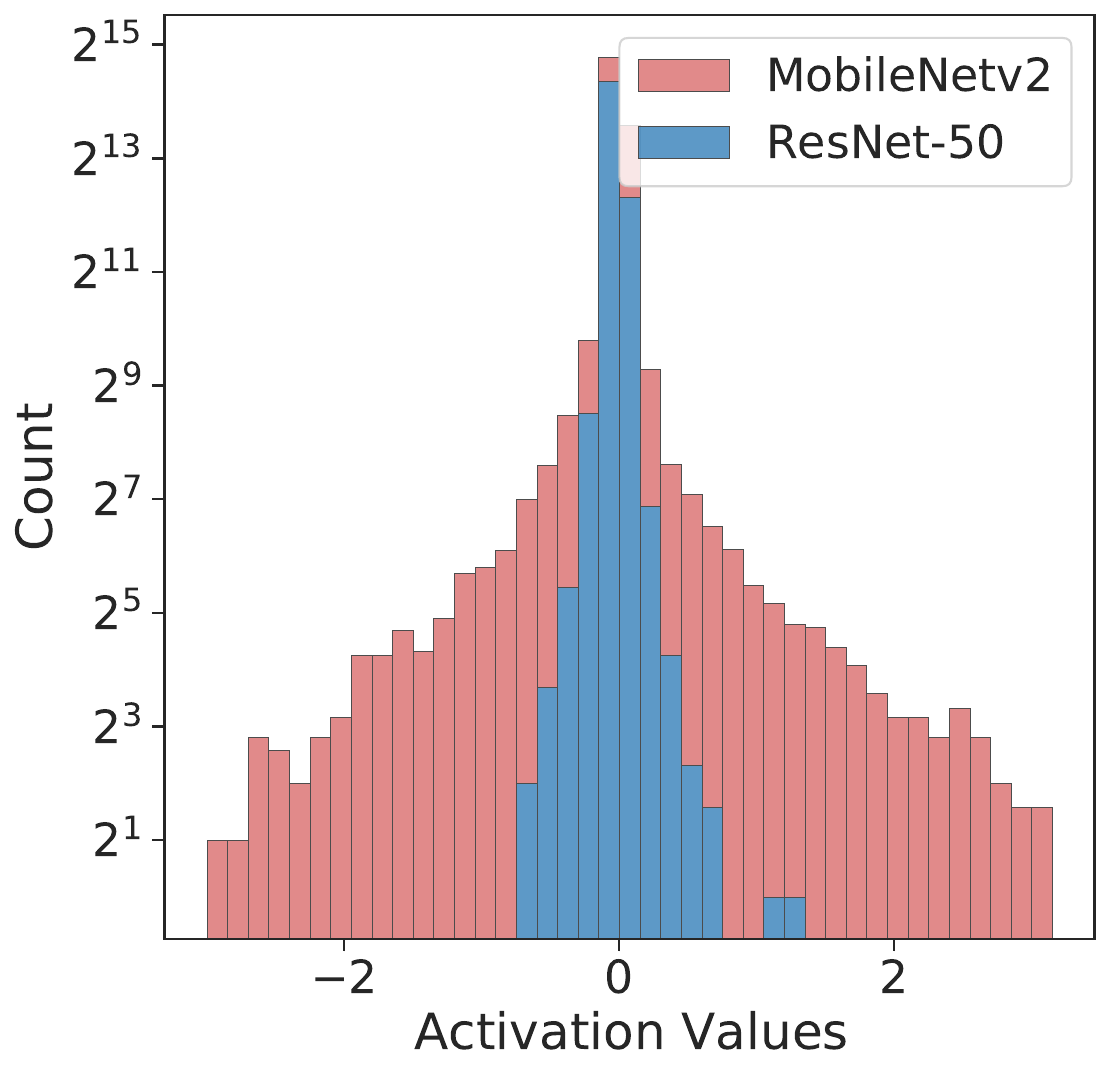}
    }
    \subfigure[]{
        \label{fig:his_5formats}
        \includegraphics[width=0.45\linewidth]{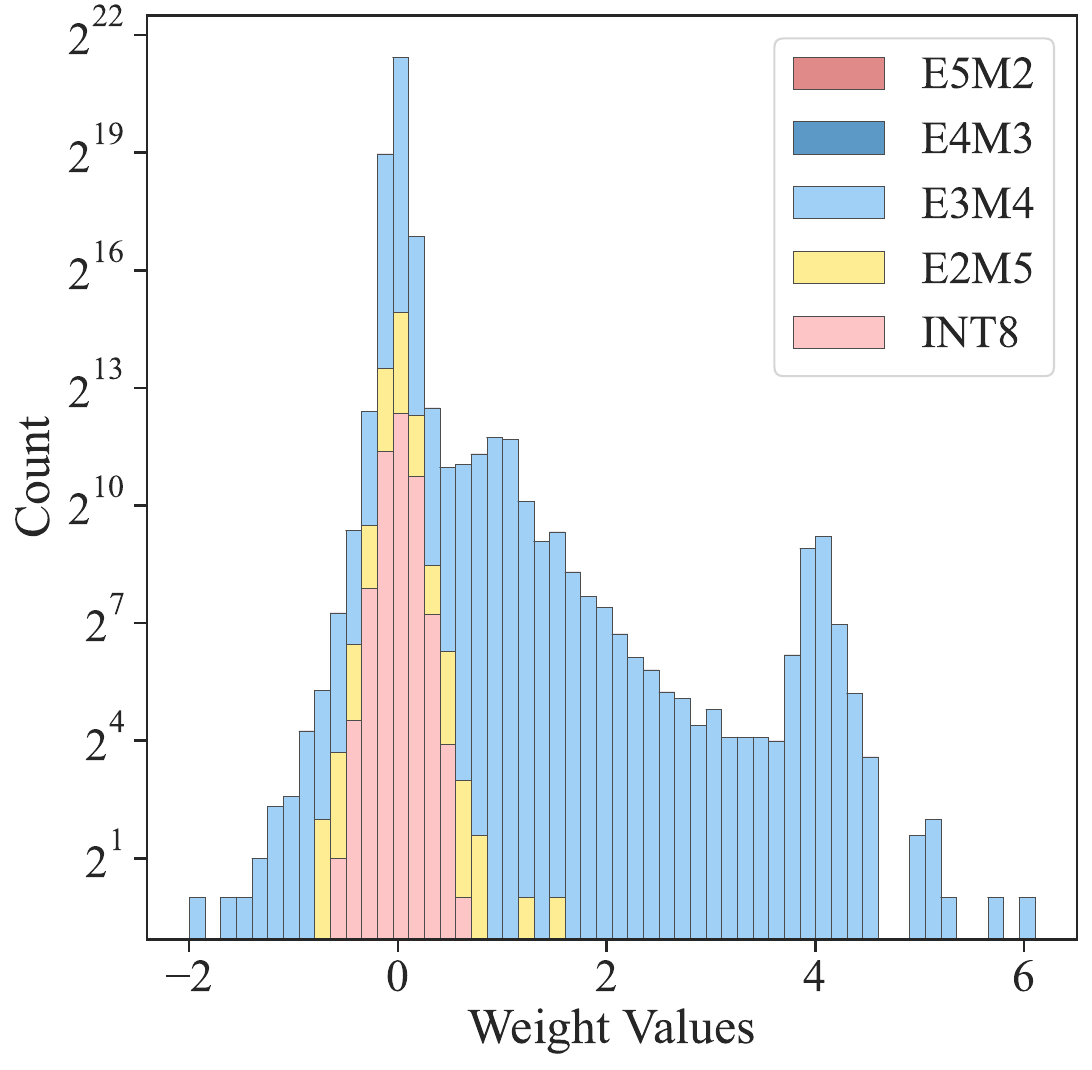}
    }
    \caption{(a) MobileNetv2 versus ResNet-50 on activation ranges. 
    (b) Accumulate a count of five formats for all values of MobileNetv2 weights.}
    \label{fig:his}
\end{figure}


To better understand the influence of format selection on data distribution, we collect the formats for all weight values in MobileNetv2, as shown in Figure~\ref{fig:his_5formats}. As expected, E3M4 dominates the format allocation, and we also observe that E2M5 is more likely to be selected for values around zero, which is resembled the behavior of INT8.




\section{Conclusion}

This paper provides a comprehensive analysis and comparison of FP8 and INT quantization methods for reducing the computational complexity of neural networks while preserving accuracy. Our experiments on a variety of tasks have demonstrated that the proposed mixed-precision quantization framework, which supports various number systems, achieves performance on par with full-precision models while providing flexible hardware design options. Our work contributes to the practical implementation of FP8 quantization and provides insights into optimal format selection for different types of neural network architectures. The proposed framework can significantly improve neural network efficiency in real-world scenarios, making it a valuable addition to the field of neural network quantization.

\bibliography{aaai24}


\clearpage
\appendix


\section{Appendix}

\subsection{Our Implementation of FP Formats}
\label{appendix:detailed_binary_formats}

Here we provide a detailed implementation of the FP8 and FP6 formats in Table~\ref{tab:detailed_fp_formats}.

\subsection{Detailed Formats Selection in Experiments}

We have investigated the selection of 8-bit and 6-bit formats for all models used in our experiments, as shown in Table~\ref{tab:baseline} and Table~\ref{tab:fp6}. 
Our findings, presented in Table~\ref{tab:detailed_format_selection}, indicate that the dominant formats are E3M4 and E3M2 in the 8-bit and 6-bit settings, respectively. We also observed that the Mixed FP8 and All Mixed approaches differ in that most of the INT8 formats are replaced by the E2M5 format, as discussed in \textbf{Section 6.4}. Similarly, for the 6-bit setting, most of the INT6 formats are replaced by the E2M3 format. Importantly, we found that switching from INT8 to E2M5 in the 8-bit setting and from INT6 to E2M3 in the 6-bit setting results in no decrease in accuracy. Therefore, increasing the number of mantissa bits in the FP number system may be a better alternative to using INT.

\subsection{Detailed Quantization Implementation}
{\bf FP Quantization Simulation.} Due to the lack of hardware support for FP8 and FP6 number systems implementation on the GPU hardware utilized in our experiments, we opted to simulate these FP formats and have made available CUDA kernels in our supplementary material to facilitate expedient experimentation. The corresponding FP8 simulation pseudocode is given in \ref{lst:fp8_simu}.

\begin{lstlisting}[caption={FP8 simulation pseudocode}, label={lst:fp8_simu}]
char _fp322fp8(x, E, M) {
    sign = (x >> (31)) & 0x1
    exp = (x >> 23) & ((1<<8)-1)
    man = x & ((1<<23) -1)
    // add exponent bias to recover the original exponent.
    exp += 1<<(E-1)-1 
    if (exp >=(1<<E)-1) 
        return max_normal(E,M)
    // round FP32 mantissa to nearest FP8 representation
    man = (man+((1<<(23-M))-1))>>(23-M)
    return sign<<7+exp<<MAN+man&((1<<MAN)-1)
}
float _fp82fp32(x, E, M) {
    if (x>>M) & ((1<<E)-1) < 1
         return  2^(1-bias)*(M1/2+M2/4+...)
    return 2^(e-bias)*(1+M1/2+M2/4+...)
}
float fp322fp8(x, E, M) {
    return _fp82fp32(_fp322fp8(x, E, M), E, M)
 }
\end{lstlisting}

{\bf Calibration.} We study the per-tensor and symmetric quantization in our experiments. The 256 calibration samples are random from the train dataset corresponding to the task. 
To quantize each layer optimally, we calculate the mean squared error (MSE) loss for each available format and select the optimal format that minimizes the MSE loss. Specifically, the format decision is given by:

\begin{equation}
    \mathop{min}\limits_{\alpha} \Vert Q^{\alpha}(\mathbf{X}) - \mathbf{X} \Vert^2, \alpha \in \mathcal{F},
\end{equation}

where $ \alpha $ is the chosen format and $ \mathcal{F} $ is the set of available formats corresponding to the bit setting. Furthermore, to consider the quantization interdependence between inputs and weights, 
we use different formats for the input and weights in convolutional and fully connected layers.
The optimal format for the layer is then determined by the MSE loss for the output of the layer, as given by:

\begin{equation}
    \mathop{min}\limits_{\alpha_{1}, \alpha_{2} \in \mathcal{F}} \Vert Q^{\alpha_{1}}(\mathbf{W}) Q^{\alpha_{2}}(\mathbf{X}) - \mathbf{W} \mathbf{X} \Vert^2,
\end{equation}

where $ \alpha_{1} $ and $ \alpha_{2} $ are the formats used for the weights $ \mathbf{W} $ and inputs $ \mathbf{X} $, respectively. By selecting the optimal format for each layer, we ensure that the formats chosen are optimized for the specific layer and avoid being trapped in the local optima of a single tensor.

\subsection{Quantitative Algorithmic Process}

The overall algorithmic flow for 8-bit quantization is displayed in the Algorithm~\ref{algorithm:quant_flow}

\begin{algorithm}
\renewcommand{\algorithmicrequire}{ \textbf{Input:}} 
\renewcommand{\algorithmicensure}{ \textbf{Output:}} 
\caption{Quantization flowchart}
\label{algorithm:quant_flow}
\begin{algorithmic}
\REQUIRE activation $ \mathbf{X} $, weight $ \mathbf{W} $, formats $\mathcal{F}$. 
\ENSURE optimal formats $ \alpha_{1} $, $\alpha_{2}$.
\STATE Compute  $\max\mathbf{\left|X\right|}$, $  \max\mathbf{\left|W\right|}$
\FOR{each $\alpha_{1} \in \mathcal{F}$}
\FOR{each $\alpha_{2} \in \mathcal{F}$}
\STATE Minimize quantization error by Eq.\ref{eq:mse} (MSE) or Eq.\ref{eq:avg_res} (Resolution-aware).
\ENDFOR
\ENDFOR
\STATE return optimal formats $ \alpha_{1} $, $\alpha_{2}$.
\end{algorithmic}
\end{algorithm}

\subsection{Discussion on Computational Cost for 8-bit Mixed Format}

The computational cost of dot product operations in FP8 and INT8 formats is dependent on hardware support. Although INT8 has been widely used due to its low computational cost, its use in some tasks can lead to significant accuracy degradation. A promising solution to this issue is mixed precision approaches that combine FP8 and INT8, which maintains reasonable accuracy while keeping the same computational costs.

It is important to note that both INT8 and FP8 have the same bandwidth, as they both use 8 bits to encode. However, the dot product operation in FP8 format has a lower computational cost than INT8 because of the difference in the type of multiplication and adder used. Specifically, INT8 uses an 8-bit multiplication and adder, while FP8 uses a 5-bit multiplier and adder.

In our study, we found that mixed format quantization with FP8 and INT8 can achieve high accuracy with lower computational costs compared to using only INT8. Moreover, our results show that using different number systems in the dot product does not lead to higher accuracy and has a higher computational cost compared to using one number system. Therefore, mixed precision approaches could be a useful technique in a variety of applications that require efficient computation without compromising accuracy. By combining the advantages of both FP8 and INT8, mixed precision approaches offer a promising direction for achieving high accuracy with lower computational costs.



\subsection{FP Format with Lower Bit}

 \begin{table*}[ht!]
    \begin{center}
    \begin{adjustbox}{max width=\textwidth}
    \begin{tabular}{l|ccccccc}
        \toprule
        \multirow{1}{*}{\bf Task (metric)} &{\bf Model} & {\bf FP32} & {\bf INT6} & {\bf Mixed FP6} & {\bf All Mixed} & {\bf Limited Mix} \\
        \midrule
        \multirow{6}{*}{Classification} & ResNet-18 & 70.28 & 35.13 & 67.99 & 68.04 & 68.08  \\
        \multirow{6}{*}{(Top-1 Acc.)} & ResNet-50 & 76.76 & 43.52 & 73.96 & 74.27 & 74.27  \\
        \multirow{6}{*}{} & MobileNetv2 & 73.26 & 25.20 & 51.86 & 53.53 & 53.53 \\
        \multirow{6}{*}{} & RegNet3.2G & 78.43 & 71.14 & 74.90 & 74.92 & 74.90 \\
        \multirow{6}{*}{} & EfficientNet-b3 & 80.19 & 0.10 & 73.73 & 73.94 & 73.93 \\
        \multirow{6}{*}{} & ViT & 83.15 & 57.71 & 82.40 &  82.47 & 82.48 \\
        \midrule
        \multirow{3}{*}{Detection} & Retina-FPN & 37.00 & 20.31 & 33.25 & 33.39 & 33.39  \\
        \multirow{3}{*}{(mAP)} & Faster-RCNN & 38.27 & 20.75 & 34.74 & 35.12 & 35.12  \\
        \multirow{3}{*}{} & YOLOX-m & 44.36 & 1.18 & 42.24 & 42.67 & 42.67  \\
        \multirow{3}{*}{} & YOLOX-L & 47.35 & 1.22 & 46.07 & 46.50 & 46.50 \\
        \midrule
        \multirow{3}{*}{Segmentation} & DeepLabv3-R50 & 79.46 & 2.58 & 70.63 & 74.43 & 74.43 \\
        \multirow{3}{*}{(mIoU)} & PSPNet-R50 & 77.46 & 1.22 & 71.15 & 74.87 & 74.87  \\
        \multirow{3}{*}{} & UNet-1024 & 67.40 & 8.55 & 60.47 & 64.23 & 64.23 \\
        \multirow{3}{*}{} & SegFormer-b0 & 75.29 & 59.11 & 72.25 & 73.72 & 73.76  \\
        \bottomrule
    \end{tabular}
    \end{adjustbox}
    \end{center}
    \caption{Precision results of various 6-bit quantization formats on classification, detection, and segmentation models. Mixed FP6 can recover precision significantly compared with INT6-only format. All Mixed version further improves performance for all models, and Limited Mix method has almost the same precision score as it. The gaps between mixing/ FP/ int at low bits are more obvious.}
    \label{tab:fp6}
\end{table*}

 We observe that 8-bit quantization is not challenging for some models like ResNet-50, among others. To further investigate the performance of INT6 and FP6, we experiment with more radical 6-bit quantization on these models. We consider the E3M2 and E2M3 in FP6 format, with the same implementation details as in FP8. Table~\ref{tab:fp6} illustrates that a reduction in bit precision aggravates the precision margin between all format combinations. In our configuration, INT6-only formats can hardly maintain precision for models such as EfficientNet-b3 and YOLOX-m, while mixed FP6 formats can recover precision scores significantly with an acceptable loss of 1-5\% in accuracy for most models. All Mixed format equipped with INT6 has a further improvement of an average of 1.17\%, which is more pronounced than the 8-bit comparison, especially in the segmentation task. This result indicates that INT6 is more complementary to FP6 formats. The \textbf{Limited Mix} method performs almost the same as the \textbf{All Mixed} methods. We find that the selections of the 6-bit format on weights and inputs between the \textbf{All Mixed} and \textbf{Limited Mix} methods are almost the same. Based on the above experiments, we conclude that the gaps among mixed format, FP, and INT at low bits are more pronounced.

\subsection{FP8 Quantization for Diffusion Model}

\begin{figure}[t]
    \centering
    \subfigure[]{
        \label{fig:diffusion_fp32}
        \includegraphics[width=0.3\linewidth]{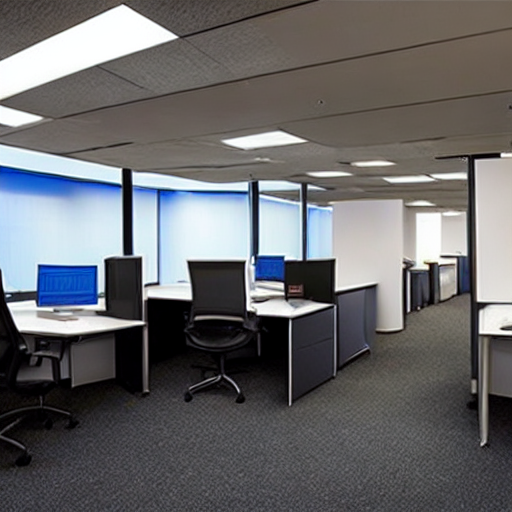}
    }
    \subfigure[]{
        \label{fig:diffusion_INT8}
        \includegraphics[width=0.3\linewidth]{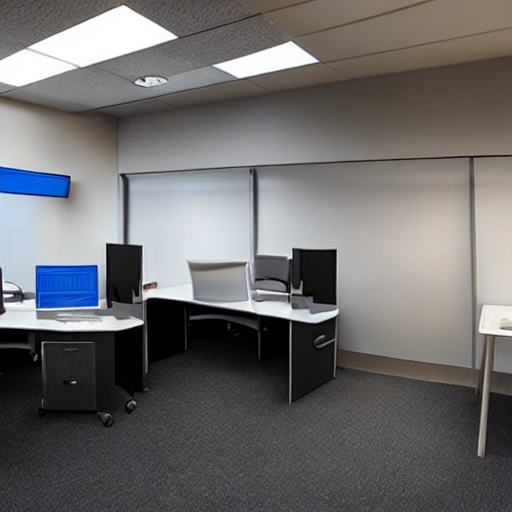}
    }
    \subfigure[]{
        \label{fig:diffusion_mix}
        \includegraphics[width=0.3\linewidth]{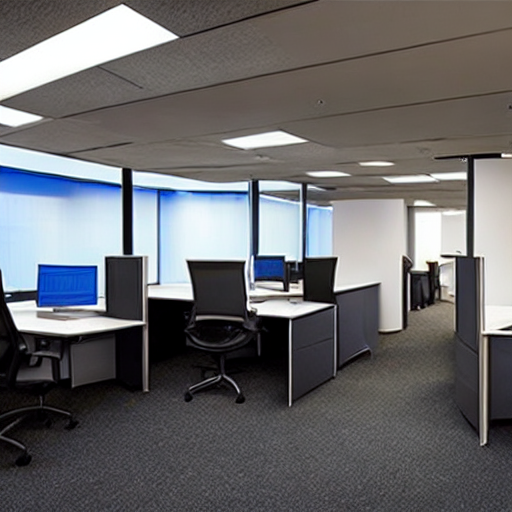}
    }
    \caption{Images generated from the FP32, INT8, and All Mixed models, respectively, are shown from left to right. The input prompt for generating the images is "An office cubicle with four different types of computers".}
    \label{fig:horse}
\end{figure}

\begin{figure*}
    \centering
    \subfigure[Full Precision]{
        \label{fig:appendix_diffusion_fp32}
        \includegraphics[width=0.3\linewidth]{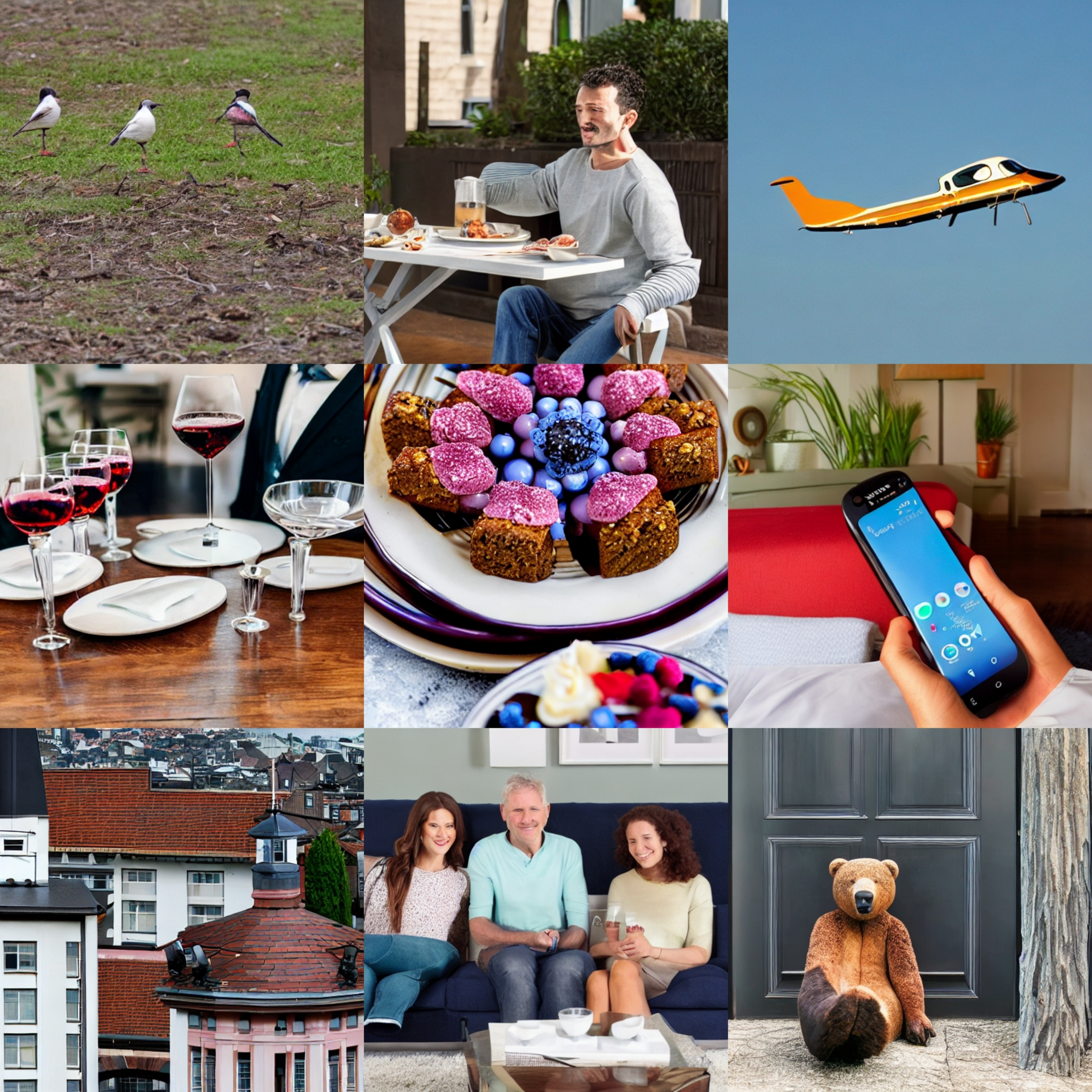}
    }
    \subfigure[INT8]{
        \label{fig:appendix_diffusion_INT8}
        \includegraphics[width=0.3\linewidth]{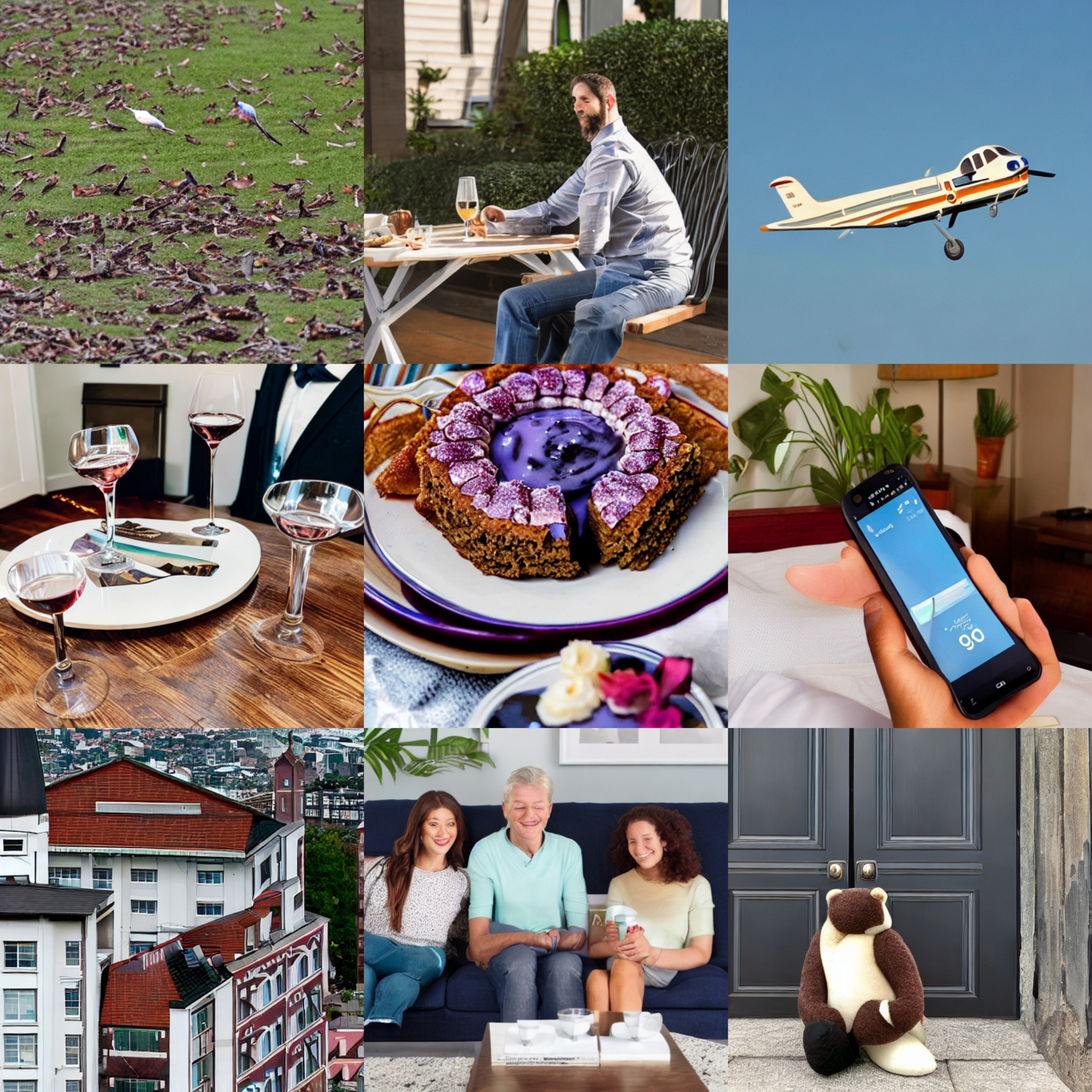}
    }
    \subfigure[FP8 (All Mixed)]{
        \label{fig:appendix_diffusion_mix}
        \includegraphics[width=0.3\linewidth]{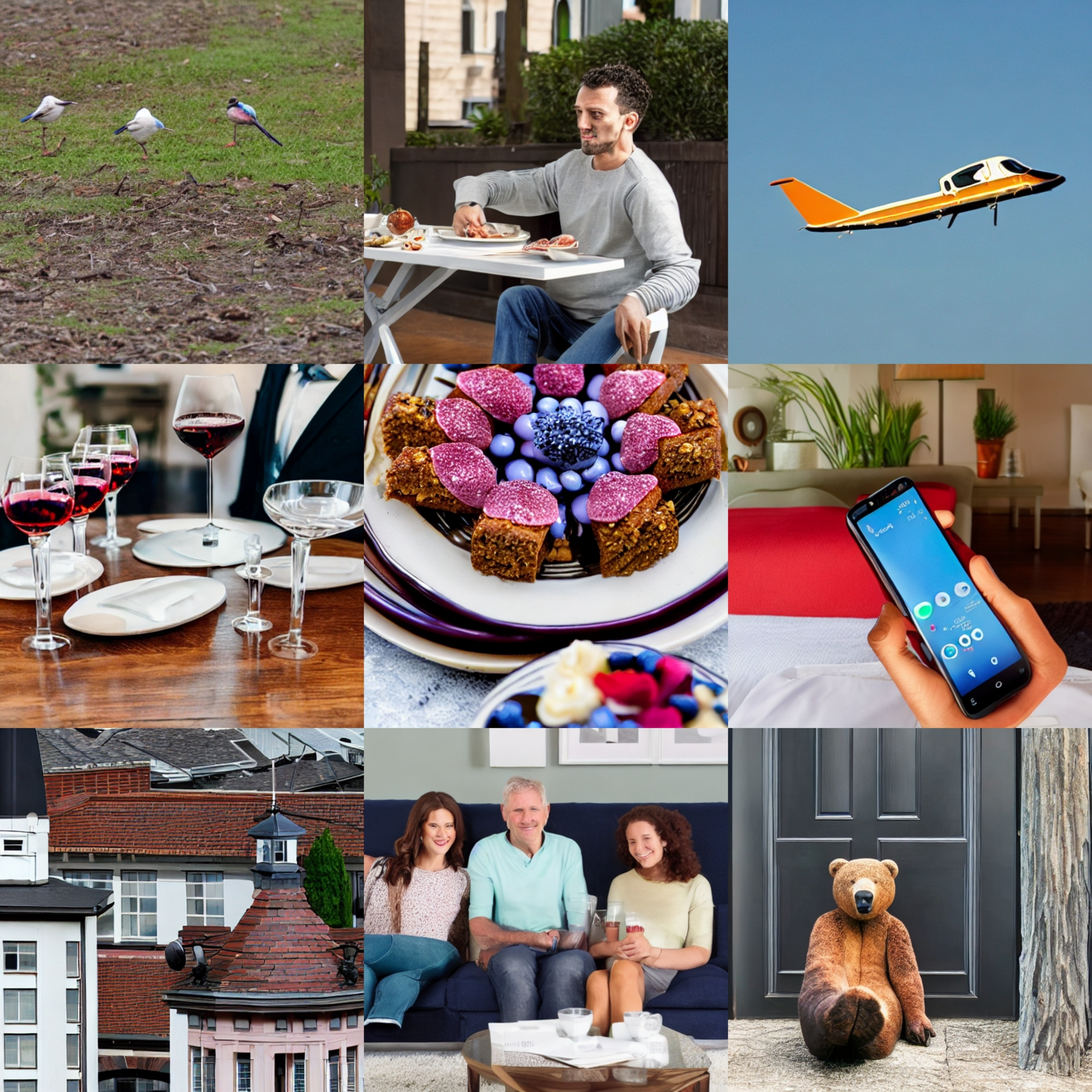}
    }
    \caption{Random samples from the FP32, INT8, and FP8 (All Mixed) models.}
    \label{fig:appendix_diffusion}
\end{figure*}

Diffusion models~\cite{ho2020denoising} present a challenge when it comes to quantization using naive PTQ methods~\cite{DBLP:journals/corr/abs-2211-15736,DBLP:journals/corr/abs-2302-04304}. To evaluate the effectiveness of our flexible FP quantization framework, we have quantized the stable-diffusion~\cite{RombachBLEO22} model for the text-to-image task. We generate a set of images, based on the caption annotations of the COCO-17 validation set, as shown in~Figure~\ref{fig:horse}. The results of our experiments, using both INT8 quantization and our flexible mixed-format quantization, are depicted in~Figure~\ref{fig:diffusion_INT8} and~Figure~\ref{fig:diffusion_mix}, respectively. Compare to the original model, which uses FP32 format and is shown in~Figure~\ref{fig:diffusion_fp32}, the image generated with INT8 format in~Figure~\ref{fig:diffusion_INT8} lacks many details, especially in the appearance of the computer. However, the mixed format image in~Figure~\ref{fig:diffusion_mix} preserves more details and comes much closer to the original model. In conclusion, our flexible mixed-format quantization framework provides a solution for the quantization of stable diffusion models. We provide more non-cherry-picked samples generated from the FP32, INT8, and FP8 (All Mixed) models. Results are shown in Figure~\ref{fig:appendix_diffusion}.

\subsection{Open Source Details}

Our code is implemented using Pytorch 1.8 and executed on a platform equipped with CUDA 11.0, python 3.6, and an Nvidia A100 GPU. More details can be found at {\it Flexible-8-bit.zip}. 

\begin{table*}[h!]
    \begin{center}
    \begin{adjustbox}{max width=\textwidth}
    \begin{tabular}{lllllll}
        \toprule
        \multirow{2}{*}{\bf Format} & \multicolumn{4}{c}{\bf 8-bit} & \multicolumn{2}{c}{\bf 6-bit} \\
        \cmidrule(l{2pt}r{2pt}){2-5}
        \cmidrule(l{2pt}r{2pt}){6-7}
         & {E5M2} & {E4M3} & {E3M4} & {E2M5} & {E3M2} & {E2M3} \\
        \midrule
        Exponent bias & 15 & 7 & 3 & 1 & 3 & 1 \\
        Inf & N/A & N/A & N/A & N/A & N/A & N/A \\
        NaN & N/A & N/A & N/A & N/A & N/A & N/A \\
        Zeros & $S.00000.00$ & $S.0000.000$ & $S.000.0000$ & $S.00.00000$ & $S.000.00$ & $S.00.000$ \\
        Max normal & $ S.11110.11_2 $ & $ S.1110.111_2 $ & $ S.110.1111_2 $ & $ S.10.11111_2 $ & $ S.110.11_2 $ & $ S.10.111_2 $ \\
        Min normal & $ S.00001.00_2 $ & $ S.0001.000_2 $ & $ S.001.0000_2 $ & $ S.01.00000_2 $ & $ S.001.00_2 $ & $ S.01.000_2 $ \\
        Max subnormal & $ S.00000.11_2 $ & $ S.0000.111_2 $ & $ S.000.1111_2 $ & $ S.00.11111_2 $ & $ S.000.11_2 $ & $ S.00.111_2 $ \\
        Min subnormal & $ S.00000.01_2 $ & $ S.0000.001_2 $ & $ S.000.0001_2 $ & $ S.00.00001_2 $ & $ S.000.01_2 $ & $ S.00.001_2 $ \\
        

        \bottomrule
    \end{tabular}
    \end{adjustbox}
    \end{center}
    \caption{The binary representation of different ours implementation for FP formats.}
    \label{tab:detailed_fp_formats}
\end{table*}

\begin{table*}[ht!]

    \begin{center}
    \begin{adjustbox}{max width=\textwidth}
    \begin{tabular}{lcccccccc}
        \toprule
        \multirow{2}{*}{\bf Task}  & \multirow{2}{*}{\bf Model} & \multicolumn{4}{c}{\bf INT8/E5M2/E4M3/E3M4/E2M5} & \multicolumn{3}{c}{\bf INT6/E3M2/E2M3} \\
        \cmidrule(l{2pt}r{2pt}){3-6}
        \cmidrule(l{2pt}r{2pt}){7-9}
        \multirow{2}{*}{} & \multirow{2}{*}{} & {Mixed FP8} & {Mixed FP8 ({\it r})} & {All Mixed} & {Limited Mix} & {Mixed FP6} & {All Mixed} & {Limited Mix} \\
        
        \midrule

        \multirow{6}{*}{Classification} & ResNet-18 & 0/0/0/75/1 & 22/12/1/41/0 & 1/0/0/75/0 & 0/0/0/75/1 & 0/75/1 & 1/75/0 & 1/75/0 \\
        \multirow{6}{*}{} & ResNet-50 & 0/0/0/156/2 & 64/34/0/60/0 & 2/0/0/156/0 & 0/0/0/156/2 & 0/156/2 & 2/156/0 & 2/156/0 \\
        \multirow{6}{*}{} & MobileNetv2 & 0/0/0/71/37 & 38/0/3/67/0 & 35/0/0/67/6 & 12/0/0/67/29 & 0/81/27 & 36/72/0 & 36/72/0 \\
        \multirow{6}{*}{} & RegNet3.2G & 0/0/0/288/1 & 53/46/9/155/0 & 2/0/0/287/0 & 2/0/0/287/0 & 0/220/69 & 98/191/0 & 0/288/1 \\
        \multirow{6}{*}{} & EfficientNet-b3 & 0/0/14/228/46 & 126/0/38/125/0 & 46/0/14/219/9 & 30/0/23/212/24 & 0/248/40 & 42/240/6 & 22/242/25 \\
        \multirow{6}{*}{} & ViT & 0/0/1/97/2 & 36/0/8/56/0 & 2/0/1/97/0 & 0/0/1/97/2 & 0/98/2 & 2/98/0 & 2/98/0 \\

        \midrule

        \multirow{4}{*}{Detection} & Retina-FPN & 0/0/0/162/10 & 53/37/1/81/0 & 10/0/0/162/0 & 0/0/0/162/10 & 0/162/10 & 10/162/0 & 10/162/0 \\
        \multirow{4}{*}{} & Faster-RCNN & 0/0/1/172/9 & 53/37/3/89/0 & 7/0/1/172/2 & 0/0/0/175/7 & 0/175/7 & 9/173/0 & 9/173/0 \\
        \multirow{4}{*}{} & YOLOX-m & 0/0/0/166/3 & 73/0/3/93/0 & 2/0/0/166/1 & 0/0/0/166/3 & 0/167/2 & 3/166/0 & 3/166/0 \\
        \multirow{4}{*}{} & YOLOX-l & 0/0/0/213/4 & 96/0/4/117/0 & 3/0/0/213/1 & 0/0/0/213/4 & 0/215/2 & 2/215/0 & 2/215/0 \\

        \midrule

        \multirow{4}{*}{Segmentation} & DeepLabv3-R50 & 0/0/0/209/5 & 77/46/2/73/0 & 1/0/0/209/4 & 1/0/0/210/3 & 0/210/4 & 4/210/0 & 4/210/0 \\
        \multirow{4}{*}{} & PSPNet-R50 & 0/0/1/200/9 & 77/44/2/71/0 & 5/0/1/200/4 & 8/0/0/200/2 & 0/202/8 & 8/202/0 & 8/202/0 \\
        \multirow{4}{*}{} & UNet-1024 & 0/0/0/55/5 & 23/15/2/20/0 & 4/0/0/55/1 & 2/0/0/56/2 & 0/56/4 & 3/56/1 & 3/56/1 \\
        \multirow{4}{*}{} & SegFormer-b0 & 0/0/1/61/23 & 38/4/3/40/0 & 26/0/1/57/1 & 14/0/1/59/11 & 0/64/21 & 24/61/0 & 24/61/0 \\

        \bottomrule
    \end{tabular}
    \end{adjustbox}
    \end{center}

    \caption{Detailed format selection for our experiments}
    \label{tab:detailed_format_selection}
\end{table*}

\end{document}